\documentclass[preprint,showpacs,preprintnumbers,amsmath,amssymb,endfloats*]{revtex4}
\usepackage{graphicx}

\begin{document}

\thispagestyle{empty}

\title{
Tests of new physics from precise measurements of the Casimir pressure 
between two gold-coated plates}

\author{R.~S.~Decca,${}^{1}$
D.~L\'{o}pez,${}^{2}$ E.~Fischbach,${}^{3}$
  G.~L.~Klimchitskaya,${}^{4}$
D.~E.~Krause,${}^{5,3}$ and V.~M.~Mostepanenko${}^{6}$}

\affiliation{${}^{1}$Department of Physics, Indiana
University-Purdue University Indianapolis, Indianapolis, Indiana 46202, USA \\
${}^{2}$Bell Laboratories, Lucent Technologies, Murray Hill, New
Jersey 07974, USA \\
${}^{3}$Department of Physics, Purdue University, West Lafayette,
Indiana 47907, USA \\
${}^{4}$North-West Technical University, Millionnaya St. 5,
St.Petersburg, 191065, Russia\\
${}^{5}$Physics Department, Wabash College, Crawfordsville,
Indiana 47933, USA \\
${}^{6}$Noncommercial Partnership ``Scientific Instruments'',
Tverskaya St. 11, Moscow, 103905, Russia}

\begin{abstract}
A micromechanical torsion oscillator has been used to strengthen
the limits on new Yukawa forces by determining the Casimir pressure
between two gold-coated plates. By significantly reducing
the random errors and obtaining the electronic parameters of
the gold coatings, we were able to conclusively
exclude the predictions of large thermal effects below 1 $\mu$m
and strengthen the constraints on Yukawa corrections to Newtonian
gravity in the interaction range from 29.5 nm to 86 nm.
\end{abstract}


\maketitle

The Casimir effect \cite{1} has attracted considerable attention
recently because it is a direct manifestation of the
existence of the quantum vacuum, and has far reaching multidisciplinary
consequences. According to Casimir's prediction there is {a}
force acting between {electrically neutral surfaces}  which
arises from the modification of the zero-point oscillations
due to {the presence of} material boundaries. (No such force {
exists} in the framework of classical electrodynamics.) The
Casimir effect finds applications in 
elementary particle physics (bag model of hadrons,
compactification of extra dimensions), gravitation and
cosmology (cosmological constant, dark matter)
 and in condensed matter physics
(layered structures, wetting processes). It is
currently being applied in nanotechnology, and to obtain constraints
on new physics beyond the standard model (references on
all of the above applications can be found in the monographs
\cite{2} and reviews \cite{6,7,8}). During the last decade
{many} experiments measuring the Casimir force between
metals, and most recently between a metal and a semiconductor,
have been performed using torsion pendulums, atomic force
microscopes, and micromechanical torsional oscillators
 \cite{9,10,13,15,19,20}.

Here we report limits of new physics obtained from an
experimental determination of the Casimir
pressure between two Au-coated plates which incorporated three
significant improvements over all previous measurements.
First, a new experimental
procedure was implemented which permitted us to {repeat the
measurements} over a wide separation region many times {in such a way
that} data were
acquired  at practically the same points for each repetition.
Second, the random experimental error was substantially reduced
compared to the systematic error, as required for precise measurements.
The total experimental relative error determined at 95\%
confidence varies from 0.19\% at the shortest separation of
$z=160$\,nm to only 9.0\% at the largest separation of $z=750$\,nm.
Third, the electronic parameters of Au films used for the extrapolation
of the tabulated optical data to low frequencies (the plasma frequency
$\omega_p$ and the relaxation parameter $\gamma$) were
determined using the measured {temperature
dependence of the film's resistivity. All these}
improvements permitted us to
obtain both precise and accurate experimental and theoretical
results over a wide {separation} range. Our results
lead to a more conclusive choice among the competing approaches to the
thermal Casimir force (see, e.g., Refs. \cite{23,24,25,26}), and to
stronger constraints on models of non-Newtonian gravity.

In this experiment,
the Casimir pressure {between} two Au coated parallel plates
was determined dynamically by means of a micromechanical torsional
oscillator consisting of a plate suspended at two opposite points
by serpentine springs, and a sphere above it attached to an optical
fiber (all details of the experimental setup are presented in
Ref.~\cite{20}). The separation between the sphere {of} radius
$R=151.3\pm 0.2\,\mu$m and the plate was varied harmonically,
$\tilde{z}(t)=z+A_z \cos (\omega_r t)$, where $\omega_r$ is the
resonant angular frequency of the oscillator under the influence
of the Casimir force from the sphere, and $A_z/z \ll 1$. (The values of
$A_z$ at each $z$ were chosen {such that the oscillator exhibited
a linear response}.) The resonant frequency $\omega_r$ is
related to the natural angular frequency $\omega_0=2 \pi
\times (713.25\pm 0.02)\,$Hz {by} \cite{15,19,20}
\begin{equation}
\omega_r^2=\omega_0^2\left[1-\frac{b^2}{I\omega_0^2}\,
\frac{\partial F(z)}{\partial z}\right], \label{eq1}
\end{equation}

\noindent where $b$ is the lever arm, $I$ is the moment of
inertia, $b^2/I=(1.2432\pm 0.0005)\,\mu\mbox{g}^{-1}$, and
$F(z)$ is the Casimir force acting between the sphere and the plate.
By measuring the frequency shift, we determine ${\partial
F}/{\partial z}$ from Eq.(1), and using the proximity force {
approximation} {(PFA)} arrive at the Casimir pressure between two
parallel plates
\begin{equation}
P(z)=-\frac{1}{2\pi R}\,\frac{\partial F(z)}{\partial z}.
\label{eq2}
\end{equation}
\noindent
Although recent results \cite{27} show
that the error in using the {PFA} is
less than $z/R$, we assume in our error analysis that the
error is $z/R$.

{This procedure for measuring the Casimir
pressure $P(z_i)$
is very similar to our previous approach \cite{20}. The main
differences {permitting us to perform measurements
 at practically the same separations
$z_i$ ($1\leq i\leq 293 $) in each of $n=33$ sets of measurements}
are a $\sim 7\%$ improvement in vibrational noise, and an
improvement in the interferometric technique used to yield the
distance $z_{meas}$ between the end of the fiber and the
stationary reference. We implemented a {two-color} fiber
interferometer similar to Ref. \cite{30b}. The use of
this interferometer yielded an error $\Delta z_{meas} = 0.2\,$nm,
and for every repetition of the Casimir pressure measurement we
were able to reposition our sample to within $\Delta z_{meas}$.}

The above experimental procedure permitted us to significantly
suppress random errors, and made it unnecessary
to resort to the cumbersome
statistical approach to data processing used in Ref.~\cite{20}.
The mean values of the Casimir pressure $\bar{P}(z_i)$ averaged
over all 33 sets of measurements are shown as black crosses in
Fig.~1a-f in different separation subintervals. (Six subfigures are
necessary to present all of the original experimental data with their
respective error bars in true scales.)
{The horizontal arms of the crosses} are equal to
twice the absolute errors, $2\Delta z=1.2$nm, determined
at 95\% confidence in the measurement
of separations between the zero roughness levels on the plate and
on the sphere (see Ref.~\cite{20} for details). The vertical arms
of the crosses are equal to twice the total absolute error,
2$\Delta P^{\rm expt}(z_i)$, determined at 95\% confidence in the
measurement of the Casimir pressure. The latter {is a function
of} separation and is shown as the solid line in Fig.~2.
{$\Delta P^{\rm expt}$} is the statistical combination
of the random experimental error (shown as the long-dashed line in
Fig.~2) obtained using the Student's $t$-distribution with
$n-1=32$ degrees of freedom, and of the systematic error (shown as
the short-dashed line in Fig.~2). The systematic error in this
experiment has two contributions: (i)
the errors in the {measurements of $\omega_r$ and  $R$}
\cite{20}, and (ii) the error from using the {PFA}.
Note that in Refs.~\cite{19,20} the latter was attributed
to theory, whereas in our current approach the theoretical
{calculation} of the Casimir pressure between two parallel plates
is independent of the {PFA}. On the other
hand, the equivalent experimental Casimir pressure  in
Eq.~(\ref{eq2}) requires the error in the {PFA}
to be attributed to the experimental systematic errors.
As is seen {in} Fig.~2, it is
the systematic error which now {dominates} the magnitude of the
total experimental error $\Delta P^{\rm expt}(z_i)$. The dominance
of the systematic error over the random error has never been
previously achieved {in  Casimir force} experiments.
The total experimental relative error ${\Delta P^{\rm
expt}}/{|\bar {P}^{\rm expt}(z)|}$ varies from 0.19\% to 9.0\%
as the separation increases from 160 to 750\,nm. (The
contribution from the use of the {PFA} to the total
experimental error varies from 0.04\% to 0.5\%, respectively.)
At short separations our experimental precision has been
improved by several times compared to our previous measurement
\cite{20}.

To compare these experimentally determined pressures with theory, one
needs to first characterize the electronic properties of the Au films used
in the experiment.  In previous experiments, these were obtained entirely
from tabulated values.
{For a more conclusive comparison of experiment with theory we
experimentally
inferred the resistivity $\rho$ of the Au films as a function of
temperature in the region from $T_1=3$K to $T_2=400$K.
Resistance
measurements were performed on Au films deposited at the same time
as the Au deposition on the oscillator, and the substrates were made
of the same material as the Si plate of the oscillator.
To do this a four probe approach
on lithographically defined Au-strips was used.
The resistivity of each sample was calculated by taking into account
its geometrical factor (each strip was
 approximately 1\,mm long, 10$\,\mu$m wide, and had the same thickness as
the film deposited on the oscillator).
The error on the resistivity of
about 3\% arises primarily from the
difficulties in accurately measuring the sample's geometry.}
The resulting data for $\rho(T)$
at $T\gg T_D/4$ (where $T_D=165$K is the Debye temperature for Au)
were fitted to a straight line.
The slope of this line was then used
to determine the magnitude of $\omega_p=8.9\,$eV.
The value of $\gamma=0.0357\,$eV at room temperature was
determined from the smooth Drude extrapolation to lower
frequencies \cite{7} of the imaginary
part of Au dielectric permittivity given by tabulated optical
data \cite{31}. Note that in previous work \cite{19,20}
the values from tables, $\omega_p=9.0\,$eV and
$\gamma=0.035\,$eV, were used.

The experimental values of the Casimir pressure in Fig.~1 were
compared with the Lifshitz theory \cite{32}
at room temperature
{$T=295\,$}K, using the calculation procedure presented
{in detail} in Ref.~\cite{20}. Note that at nonzero {
Matsubara} frequencies the use of the reflection coefficients
expressed in terms of dielectric permittivity \cite{23,24,25},
or in terms of Leontovich impedance \cite{26}, leads to
negligibly small differences in the final results.
However, at zero
frequency this is not the case. In terms of the Leontovich
impedance, the reflection coefficients for the transverse magnetic
(TM) and transverse electric (TE) modes are given by \cite{26}
\begin{equation}
r_{\rm TM}(0,k_{\bot})=1,\qquad r_{\rm
TE}(0,k_{\bot})=\frac{\omega_p-ck_{\bot}}{\omega_p+ck_{\bot}},
\label{eq3}
\end{equation}
\noindent
where $k_{\bot}$ is the magnitude of the wave vector in
the plane of the plates. In contrast, when the reflection
coefficients are expressed in terms of the dielectric
permittivity of the Drude model one finds \cite{23,33}
\begin{equation}
r_{\rm TM}(0,k_{\bot})=1,\qquad r_{\rm TE}(0,k_{\bot})=0.
\label{eq4}
\end{equation}
\noindent
Importantly, Eq.~(3) predicts small thermal effects at
$z\leq 1\,\mu$m in qualitative agreement with the case of ideal
metals. However, Eq.~(4) results in {relatively} large
thermal {effects} up to 16\% of the Casimir pressure at $z\leq
1\,\mu$m, and an 11\% effect at $z=750\,$nm.
Further discussions on the applicability of
Eqs.~(3) and (4) can be found in Refs.~\cite{34,35}.

The theoretical Casimir pressures were computed by using both
Eq.~(3) and Eq.~(4) (the Leontovich impedance approach and the
Drude model approach, respectively). The surface roughness with
{variances} equal to 3.6\,nm on the plate and 1.9\,nm on the
sphere {was} taken into account by the geometrical averaging
{method} \cite{20}. Roughness contributes
a correction of only 0.5\% of the
Casimir pressure at $z=160$ nm, and its contribution decreases with
the increase of separation. The contributions of
diffraction-type and correlation effects in the roughness
correction, which are {not considered} in the geometrical
averaging, were shown to be negligible \cite{20}. The theoretical
Casimir pressures taking into account the surface roughness are shown
in Fig.~1a-f by the light-gray bands (the Leontovich impedance
approach), and by the dark-gray bands (the Drude model approach).
The width of the bands in the vertical direction is equal to twice the
total theoretical error 2$\Delta P^{\rm theor}(z)$,{
determined} at 95\% confidence. This error arises from the
variation of the tabulated optical data and extrapolation parameters
and results in a
relative theoretical error ${\Delta P^{\rm theor}(z)}/{|P^{\rm
theor}(z)|}$ equal to 0.5\%
\cite{20}. {Other factors, such as patch potentials or spatial
nonlocality,} were shown to be negligible (see Ref.~\cite{20} for
more details). In our present approach to the comparison of
experiment with theory, the theoretical pressures are not computed
at the experimental points but rather within the measurement range from
160 to 750\,nm. Because of this, the error in the measurement of
separation $\Delta z$ is irrelevant to theory and should be included
in the analysis of experimental errors.
As {seen in} Fig.~1a-f, the Leontovich
impedance approach is consistent with data over the entire
measurement range.
We note that by using our new values of $\omega_p$ and $\gamma$,
better agreement is achieved between the data and
the impedance approach within the
separation region from 200 to 400\,nm. With the values previously used,
differences $(P^{\rm theor}-P^{\rm expt})$ were changing non-monotonically
with decreasing $z$, but with the values determined here
the expected monotonic dependence on $z$ is observed.
We emphasize, however, that the variations in $P^{\rm theor}$ using both
sets of $\omega_p$ and $\gamma$ lie within the limit of theoretical error.
In contrast, the Drude model approach, leading to relatively
large thermal effects at $z\leq 1\,\mu$m, is excluded
experimentally at 95\% confidence over the entire measurement range
from 160 to 750 nm. If we choose a confidence level of
99\% or 99.9\%, the Drude model approach is excluded by the
experimental data from 160 to 700\,nm, and from 210 to 620\,nm,
respectively.
The increase in the experimental precision achieved in this work, along
with the use of more accurate values of $\omega_p$ and
$\gamma$, allowed us to widen
the interval in which the Drude model approach is excluded at 99\%
confidence (in Ref.~\cite{20} from 300 to 500\,nm), and
to demonstrate for the first
time the exclusion of this model at the 99.9\% confidence level.
Note that our results do not compromise applications of the Drude model
other than in the thermal Casimir effect. The Casimir pressure computed by
the Lifshitz formula at $T=0$ is also consistent with our data.
The small thermal effects predicted {in Refs.~\cite{24,25,26}}
are currently experimentally inaccessible.

Our data can be used to impose stronger constraints on a 
Yukawa correction to the Newtonian gravitational potential
predicted from the exchange of light hypothetical bosons
\cite{36}, and from extra-dimensional physics with {a low
unification} scale \cite{37}. The potential energy between two
point masses $m_1$ and $m_2$ a distance $r$ apart arising from
Newton's law {with a} Yukawa correction {is} given by
\begin{equation}
V(r)=-\frac{Gm_1m_2}{r}\left(1+\alpha e^{-r/\lambda}\right),
\label{eq5}
\end{equation}
\noindent
where $G$ is the gravitational constant, $\alpha$
characterizes the strength of the Yukawa force, and $\lambda$ is
its interaction range.
The new constraints on $\alpha,\>\lambda$ are obtained
using the same procedure as in Ref.~\cite{20}, and
are shown in Fig.~3 by line 1. In the same figure the constraints
from earlier experiments are also shown. As {seen in} Fig.~3,
the new constraints are the strongest in the interaction region
29.5\,nm$\leq\lambda\leq$86\,nm with the largest improvement by a
factor of 3 at $\lambda\approx 40\,$nm.

To summarize, our new results extend the range of separations in
which we achieve a small experimental error. In addition, we have
for the first time used the experimentally determined values of
the electronic parameters of the samples prepared under the same
conditions as the Au deposition on the oscillator when comparing
to theory. This has permitted us to
conclusively exclude the approach to the thermal Casimir force
predicting large thermal effects at separations below $1\mu$m, and
to {strengthen} constraints on non-Newtonian gravity. The
experimental procedures we have developed
will find applications not only in
different fields of fundamental physics but also in nanomechanical
devices driven by the Casimir force.

The work of E.F. was supported in part by DOE under the Contract
No. DE-AC02-76ER071428.
R.S.D. acknowledges NSF support through grant CCF-0508239.
G.L.K. and V.M.M. were partially supported
by DFG grant 436 RUS 113/789/0-2.



\begin{figure*}
\vspace*{-2cm}
\includegraphics{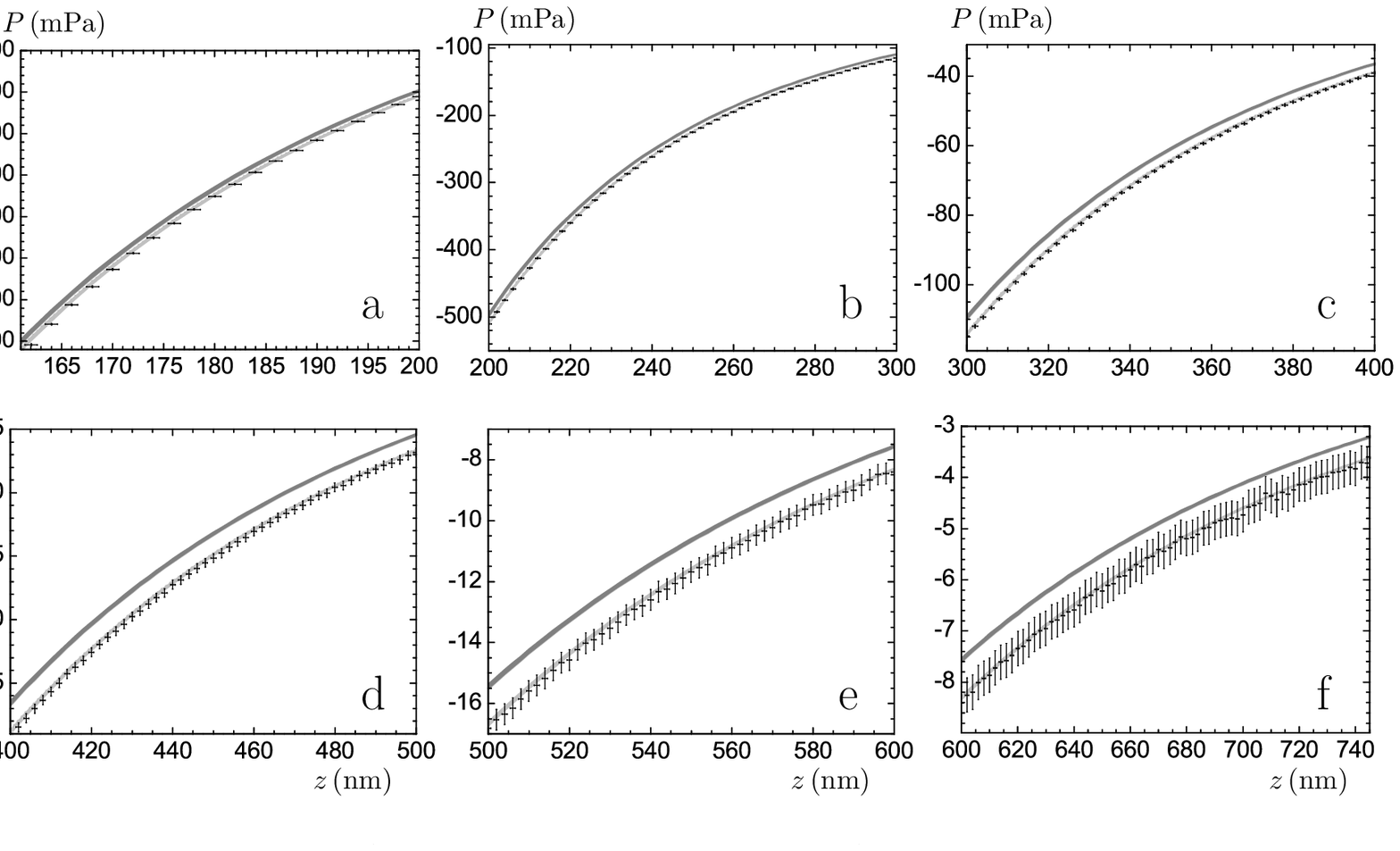}
\vspace*{-11cm}
\caption{Experimental data for the Casimir pressure as a function
of separation $z$.
Absolute errors are shown by black crosses in different separation
regions (a--f). The light- and dark-gray bands represent the
theoretical predictions of the impedance and Drude model
approaches, respectively. The vertical width of the bands is equal
to the theoretical error, and all crosses are shown in true scale. }
\end{figure*}


\begin{figure*}
\vspace*{-7cm}
\includegraphics{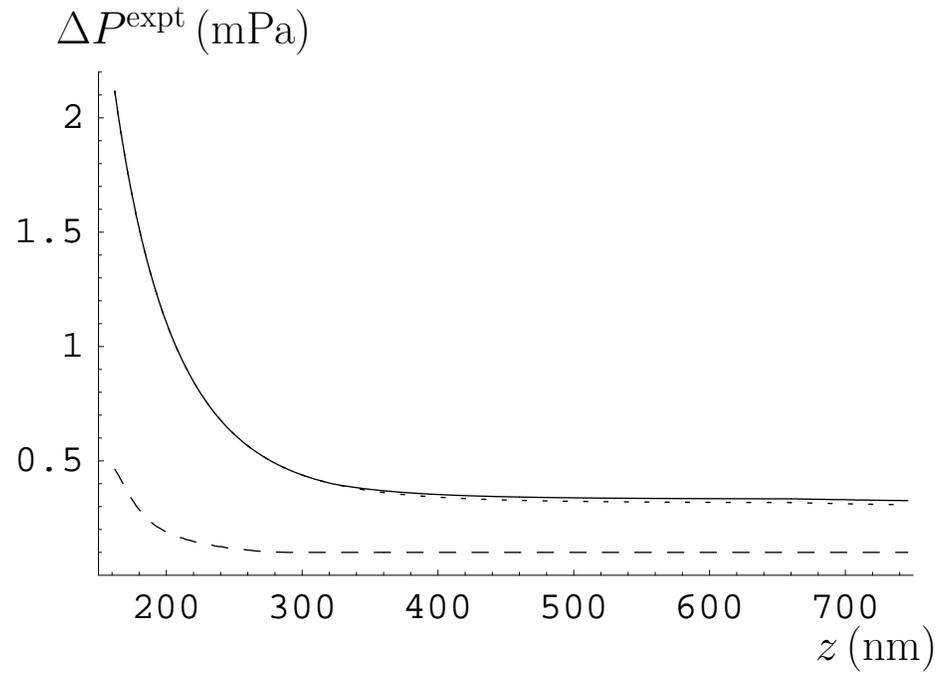}
\vspace*{-10cm} \caption{The total absolute experimental error of
the Casimir pressure measurements (solid line), random error
(long-dashed line), and systematic error (short-dashed line) as
functions of separation.}
\end{figure*}


\begin{figure*}
\vspace*{-7cm}
\includegraphics{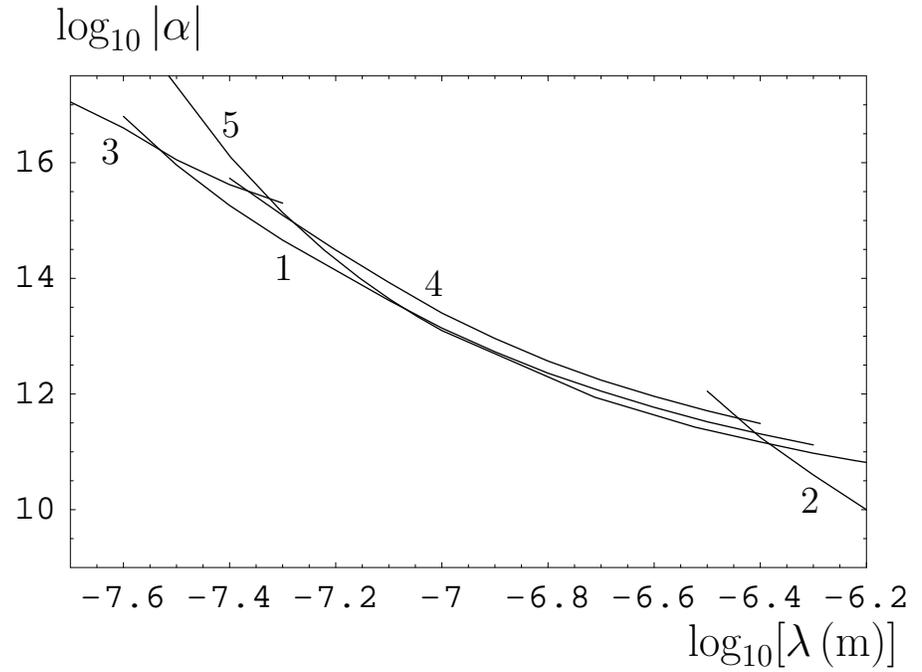}
\vspace*{-10cm}
\caption{Constraints on the parameters of the
Yukawa interaction obtained in this paper (line 1), and from the
experiments of Refs. \cite{9,13,20,38} (lines 2, 3, 4, and 5,
respectively).  The region of $(\lambda,\alpha)$ plane above each
line is excluded, and that below is allowed.}
\end{figure*}


\end{document}